\documentclass[12pt]{article}
\usepackage{graphicx}
\usepackage{hyperref}
\usepackage{amsmath}
\usepackage{amssymb} 
\usepackage{color}
\definecolor{mypurple}{rgb}{0.71,0.02,1}

\def\be{\begin{equation}}
\def\ee{\end{equation}}
\def\bea{\begin{eqnarray}}
\def\eea{\end{eqnarray}}
\def\bi{\begin{itemize}}
\def\ei{\end{itemize}}
\def\bs{\begin{slide}}
\def\es{\end{slide}}
\def\dd{\mathrm{d}}
\date{}
\title{Non-uniqueness of the Dirac theory\\ in a curved spacetime}
\author{
Mayeul Arminjon\,$^1$ and Frank Reifler\,$^2$\\
$^1$ \small\it CNRS (Section of Theoretical Physics),\\
\small\it Laboratory ``Soils, Solids, Structures, Risks'', Grenoble, France\\
\small e-mail: arminjon@hmg.inpg.fr\\
\small\it $^2$ Lockheed Martin Corporation, MS2 Division,
Moorestown, New Jersey, USA
} 

\begin{document}
\maketitle

\begin{abstract}
\noindent We summarize a recent work on the subject title. The Dirac equation in a curved spacetime depends on a field of coefficients (essentially the Dirac matrices), for which a continuum of different choices are possible. We study the conditions under which a change of the coefficient fields leads to an equivalent Hamiltonian operator H, or to an equivalent energy operator E. In this paper, we focus on the standard version of the gravitational Dirac equation, but the non-uniqueness applies also to our alternative versions. We find that the changes which lead to an equivalent operator H, or respectively to an equivalent operator E, are determined by initial data, or respectively have to make some point-dependent antihermitian matrix vanish. Thus, the vast majority of the possible coefficient changes lead neither to an equivalent operator H, nor to an equivalent operator E, whence a lack of uniqueness. We show that even the Dirac energy spectrum is not unique. 
\end{abstract}

\section{Introduction}
This paper summarizes a recent work \cite{A43} on the subject title. Thus, rather than to give detailed arguments and to present all relevant results, our aim here is to make the origin of this surprising result more easily visible. 

Quantum effects in the classical gravitational field {\it are observed,} e.g. on neutrons \cite{COW1975,WernerStaudenmannColella1979,Nesvizhevsky2002}, which are spin $\frac{1}{2}$ particles. This motivates work on the curved spacetime Dirac equation. The standard version of the latter (see e.g. Refs. \cite{BrillWheeler1957+Corr,ChapmanLeiter1977}) is due to Fock and to Weyl, and will be referred to as {\it the Dirac-Fock-Weyl} ({\bf DFW}) {\it equation}. On the other hand, two alternative Dirac equations in a curved spacetime were derived recently \cite{A39}, by applying directly the ``classical-quantum correspondence", i.e., the correspondence between a classical Hamiltonian and a quantum wave operator. Thus, together with the standard (DFW) version, we have three distinct versions of the Dirac equation in a curved spacetime. The basic quantum mechanics was studied for each of those three equations \cite{A42} (see Ref. \cite{B30} for a summary):
\bi

\item the Dirac probability current was defined for any possible field $(\gamma ^\mu )$ of Dirac matrices. We derived the condition which has to be satisfied by $(\gamma ^\mu )$ in order that the current conservation be true for any solution of the Dirac equation (for DFW, this condition is always satisfied);
\item the precise form of the relevant scalar product was shown to be imposed by the axioms of quantum mechanics (it turns out that, for DFW, this is the form used by Leclerc \cite{Leclerc2006});
\item the Hamiltonian operator was written and the condition of its hermiticity w.r.t. the relevant scalar product was derived in a general setting (for DFW, this condition coincides, in the particular case envisaged by Leclerc \cite{Leclerc2006}, with that derived by him). 

\ei
That foregoing work showed, in particular, that the hermiticity of the Hamiltonian is not stable under all admissible changes of the field of Dirac matrices. This implies that there is a non-uniqueness problem for the curved-spacetime Dirac equation.\\

The {\it aim of the present work \cite{A43},} therefore, was to study the (non-)uniqueness of the Hamiltonian and energy operators, including the (non-)uniqueness of the energy spectrum. The qualitative conclusions are the same for the three versions, i.e., the non-uniqueness applies to the alternative equations of Ref. \cite{A39}, too \cite{A43}. Thus, finding this non-uniqueness was disappointing for us as well. In this paper, we will focus on the {\it standard equation} (DFW) and the non-uniqueness results for it, because these results could be more directly of interest to many physicists.

  \section {Three Dirac equations in a curved spacetime}

The three versions of the gravitational Dirac equation have the same form: 
\be \label{Dirac-general}
\gamma ^\mu D_\mu\psi=-im\psi,\\
\ee
\noindent where $\gamma ^\mu =\gamma ^\mu (X)$ ($\mu =0,...,3$) is a field of $4\times 4$ complex matrices, depending on the point $X\in \mathrm{V}$ in the curved spacetime $\ (\mathrm{V},g_{\mu \nu})$, such that
\be \label{Clifford}
\gamma ^\mu \gamma ^\nu + \gamma ^\nu \gamma ^\mu = 2g^{\mu \nu}\,{\bf 1}_4, \quad \mu ,\nu \in \{0,...,3\} 
\ee
[here $(g^{\mu \nu})(X)$ is the inverse of the matrix $(g_{\mu \nu})(X)$ and ${\bf 1}_4\equiv \mathrm{diag}(1,1,1,1)$];\\ where $\psi$ is a {\it bispinor} field for the standard equation (DFW), but is a {\it 4-vector} field for the two  alternative equations, based on the {\it tensor representation of the Dirac fields} ({\bf TRD});\\
\noindent and where {$D_\mu $} is a covariant derivative, associated with a specific {\it connection}. For DFW, this is the ``spin connection", which depends on the $(\gamma ^\mu)$ field \cite{BrillWheeler1957+Corr,ChapmanLeiter1977}.

 \section{Definition of the field of Dirac matrices}

{\it For DFW,} one defines 
\be\label{gamma from tetrad}
\gamma ^\mu(X)= a^\mu_{\ \,\alpha}(X) \ \gamma ^{ \sharp \alpha}, 
\ee
with  {$u_\alpha =a^\mu_{\ \,\alpha}(X)\, \partial _\mu$}  an orthonormal tetrad field ($\partial _\mu$ is the natural basis associated with the coordinates $x^\mu $) and {$(\gamma ^{\sharp \alpha}) $} a set of ``flat" Dirac matrices, i.e., a constant solution of the anticommutation relation (\ref{Clifford}) with $(g^{\mu \nu })=(\eta ^{\mu \nu })\equiv \mathrm{diag}(1,-1,-1,-1)$. One should be able to use {\it any} possible choice of {$(\gamma ^{\sharp \alpha} )$}. One should study the influence of both choices:  {$(\gamma ^{\sharp \alpha}) $} and ${(u_\alpha)}$. 

{\it For TRD,} a tetrad field can also be used, though other possibilities also exist, e.g. parallel transport \cite{A39}.\\

In order to be able to use any possible field {$(\gamma  ^\mu )$} of Dirac matrices, we must have recourse to the {\it hermitizing matrix} {$A$} of Bargmann and Pauli \cite{Pauli1933}. This is a {$\ 4 \times 4$} complex matrix such that
\be\label{hermitizing-A}
{A^\dagger = A, \qquad (A\gamma ^\mu )^\dagger = A\gamma ^\mu \quad \mu =0, ...,3},
\ee
with {$M^\dagger\equiv M^{*\,T}$} = Hermitian conjugate of matrix {$M$}. We proved \cite{A40} the existence of {$A$}, and also that of ${B}$: the latter is a positive-definite hermitizing matrix for the ${\alpha^\mu}$ matrices which are defined by Eq. (\ref{alpha}) below. In general, $A=A(X)$ is also a field. However, when we use the definition from a tetrad field (\ref{gamma from tetrad}), we may take \cite{A42}
\be\label{A = Asharp}
A\equiv A^\sharp,
\ee
where $A^\sharp$ is a (constant) hermitizing matrix for the constant set $(\gamma ^{\sharp \alpha})  $. Moreover, in practice for DFW, one considers only sets {$(\gamma ^{\sharp \alpha})  $} for which $\gamma ^{\sharp 0}$ is a hermitizing matrix. (This includes the usual sets: Dirac's, ``chiral", Majorana's, and any set deduced therefrom by a unitary transformation \cite{Pal2007}.) Thus, in practice,
\be\label{A = gamma sharp 0}
 \underline{A= \gamma ^{\sharp 0}\ \mathrm{for\ DFW}.}\\
\ee

  \section{Local similarities}

In a curved spacetime {$(\mathrm{V},g_{\mu \nu })$}, the Dirac matrices {$\gamma ^\mu$} and the hermitizing matrix {$A$} are fields: they depend on {$X \in \mathrm{V}$}. If one changes from one field {$(\gamma ^\mu)$} to another one {$(\tilde {\gamma }^\mu)$}, also satisfying the anticommutation relation (\ref{Clifford}) and with the {\it same} metric $g_{\mu \nu}$, then the new field obtains by a {\it local similarity transformation} \cite{Pauli1933}, which is unique up to a complex factor \cite{A40}:
\be \label{similarity-gamma}
{\exists S=S(X) \in {\sf GL}(4,{\sf C}):\qquad \tilde{\gamma} ^\mu(X) =  S^{-1}\gamma ^\mu(X) S, \quad \mu =0,...,3}.
\ee
For the standard Dirac equation (DFW), the similarities are restricted to the spin group {${\sf Spin(1,3)}$}, i.e., they result from a change of the tetrad field by a (proper) local Lorentz transform $L(X)\in {\sf SO(1,3)}$, depending arbitrarily on $X \in \mathrm{V}$ \cite{BrillWheeler1957+Corr}. The corresponding local similarity transformation $S(X) \in {\sf Spin(1,3)}$ is deduced \cite{A42} from $L(X)$ through the spinor representation (defined only up to a sign), $L\mapsto S=\pm {\sf S}(L)$.


  \section{\bf The general Dirac Hamiltonian} 

Rewriting the Dirac equation (\ref{Dirac-general}) in the ``Schr\"odinger" form:
\be \label{Schrodinger-general}
{i \frac{\partial \psi }{\partial t}= \mathrm{H}\psi,\qquad (t\equiv x^0)},
\ee
gives the Hamiltonian operator:
\be \label{Hamilton-Dirac-general}
{\mathrm{H} \equiv  m\alpha  ^0 -i\alpha ^j D _j -i(D_0-\partial_0) }, 
\ee
with
\be \label{alpha}
{\alpha ^0 \equiv \gamma ^0/g^{00}}, \qquad {\alpha ^j \equiv \gamma ^0\gamma ^j/g^{00} \quad (j=1,2,3)}.
\ee
One should realize that, in a given spacetime $(\mathrm{V},g_{\mu \nu })$ and with a given field of Dirac matrices $(\gamma ^\mu)$, this operator still depends on the coordinate system, or, more exactly, on the {\it reference frame} \cite{A42}---the latter being understood here as an equivalence class F of local coordinate systems (charts) on the spacetime V, modulo the purely spatial transformations
\be\label{purely-spatial-change}
x'^0=x^0,\ x'^j=f^j((x^k)).
\ee
In the present work, we fix the reference frame F and even the chart (the latter being, however, an arbitrary one), and we study the dependence of the Hamiltonian and energy operators on the field $(\gamma ^\mu)$.

\section{Scalar product and equivalent operators}

The Hilbert scalar product is fixed by the following result \cite{A42}:
\paragraph{Theorem.}\label{Theorem5} {\it A} necessary {\it condition for the scalar product of time-independent wave functions to be time independent and for the Hamiltonian $\mathrm{H}$ to be a Hermitian operator, is that the scalar product should be}
\be \label{Hermitian-sigma=1-g}
(\psi  \mid \varphi  ) \equiv \int \psi^\dagger A\gamma ^0  \varphi \ \ \sqrt{-g}\ \dd^ 3{\bf x}.
\ee

For DFW, we have in practice $A=\gamma^{\sharp 0}$, Eq. (\ref{A = gamma sharp 0}). Then (\ref{Hermitian-sigma=1-g}) is the scalar product used by Leclerc \cite{Leclerc2006}.

With each of any two possible ``coefficient fields": $(\gamma ^\mu ,A)$ and $(\tilde {\gamma}^\mu,\tilde {A})$, corresponds thus a unique scalar product. [For DFW, we have always $\tilde {A}=A=A^\sharp$, Eq. (\ref{A = Asharp}), so the coefficient fields reduce to the field of Dirac matrices $\gamma ^\mu $.] These two scalar products are {\it isometrically equivalent} through the mapping $\psi \mapsto \tilde { \psi } \equiv S^{-1}\psi $: one shows easily \cite{A43} that
\be \label{tilde=isometry}
(\tilde { \psi }  \,\tilde { \mid} \,\tilde { \varphi }  ) =(\psi \mid \varphi ).
\ee
Moreover, H is fully determined by the set of the products $(\mathrm{H} \, \psi \mid    \varphi  )$, for $\psi ,\varphi \in \mathcal{D} \equiv \mathrm{Dom}(\mathrm{H})$. The same is true for the Hamiltonian $\tilde {\mathrm{H}}$ after the similarity (\ref{similarity-gamma}), $\tilde {\mathrm{H}}$ being obtained by substituting $\tilde {\gamma}^\mu$ for $\gamma^\mu$ in Eq. (\ref{alpha}), and by using the covariant derivatives $\tilde{D}_\mu \equiv \partial_\mu + \tilde{\Gamma}_\mu$ in (\ref{Hamilton-Dirac-general}), with $\tilde{\Gamma}_\mu$ the new connection matrices [see Eq. (\ref{Gamma-tilde-psitilde=S^-1 psi}) below]. It follows \cite{A43} that, in order that H and $\tilde {\mathrm{H}}$ be equivalent operators, it is necessary and sufficient that
\be \label{similarity-invariance-H}
\tilde{\mathrm{H} }  =  S^{-1}\,\mathrm{H}\, S,
\ee
where $S$ is the relevant similarity in Eq. (\ref{similarity-gamma}). This, of course, is no surprise. Of course also, the same condition defines the equivalence of the energy operators E and $\tilde{\mathrm{E} }$, E being defined by Eq. (\ref{H^s = E}) below.

  \section{Invariance condition of the Hamiltonian under a local similarity (DFW)}

When does a local similarity {$S(X)$}, applied to the field of Dirac matrices {$(\gamma ^\mu)$}, leave the Hamiltonian operator $\mathrm{H}$ [equation (\ref{Hamilton-Dirac-general})] invariant? I.e., when do we have Eq. (\ref{similarity-invariance-H})?

The Hamiltonian (\ref{Hamilton-Dirac-general}) involves the connection matrices, {$\Gamma _\mu\equiv D_\mu -\partial _\mu $}. For DFW, these matrices change after a similarity according to \cite{ChapmanLeiter1977}:
\be\label{Gamma-tilde-psitilde=S^-1 psi}
{\tilde{\Gamma }_\mu = S^{-1}\Gamma _\mu S+ S^{-1}(\partial _\mu S)}.
\ee
A straightforward calculation shows then that, for DFW, we have (\ref{similarity-invariance-H}) iff {$S(X)$} is time-independent, 
\be
\partial_0 S=0.
\ee
But, in the general case, we have {$g_{\mu \nu ,0} \ne 0$}, hence any possible field {$(\gamma ^\mu)$} depends on {$\ t\equiv x^0 $}. Thus, there is no way of finding a class of fields {$(\gamma ^\mu)$} exchanging with  {$\partial_0 S=0$}. I.e.: {\it The Dirac Hamiltonian is not unique} \cite{A43}.\\

  \section{Invariance condition of the energy operator (DFW)}

When the Hamiltonian {$\mathrm{H}$} is not Hermitian, one should use the Hermitian part of {$\mathrm{H}$}\,:
\be\label{H^s = E}
{\mathrm{E} =  \mathrm{H}+\frac{i}{2\sqrt{-g}}\,B^{-1}\,\partial _0 \left(\sqrt{-g}\,B\right) = \frac{1}{2}\left(\mathrm{H}+\mathrm{H}^\ddagger \right)}, \ {B\equiv A\gamma ^0},
\ee
where the Hermitian conjugate operator $\mathrm{H}^\ddagger$ is w.r.t. the unique scalar product (\ref{Hermitian-sigma=1-g}). This Hermitian-symmetrized operator E coincides, in the particular case envisaged by Leclerc \cite{Leclerc2006}, with what is called the {\it energy operator,} which is derived from the field Lagrangian. In the general case (thus for TRD and for DFW as well, and with any possible $(\gamma ^\mu )$ field), the Dirac equation (\ref{Dirac-general}) also derives from a Lagrangian \cite{A43}. We can show generally that the expected value of the energy operator E equals the classical field energy as derived from that Lagrangian.

Again a straightforward calculation gives the invariance condition of {$\mathrm{E}$} (for DFW):
\be\label{SEtilde=ES DFW}
{B(\partial _0S)S^{-1}-\left[B(\partial _0S)S^{-1}\right]^\dagger \equiv 2 \left[B(\partial _0S)S^{-1}\right]^a = 0}.
\ee
Only very particular local similarities {$S(X)$} do verify (\ref{SEtilde=ES DFW}). Thus, there is a serious non-uniqueness problem for DFW (and for the alternative, TRD equations as well). Could even the {\it spectrum} of {$\mathrm{E}$} be non-unique? In what follows, we investigate this question.

  \section{Explicit expression of the energy operator (DFW)}

The general expression of the change of ${\mathrm{E}}$ after a local similarity $S(X)$ is easily found to be \cite{A43}:
\be\label{SEtilde-ES-DFW-1}
{\delta \mathrm{E}\equiv S\tilde{\mathrm{E}}S^{-1} -\mathrm{E} =-iB^{-1} \left[B(\partial _0S)S^{-1}\right]^a}.
\ee
In very general coordinates, the matrix $(a^\mu _{\ \,\alpha} )$ of that tetrad which defines the starting $(\gamma ^\mu )$ field may be chosen to satisfy $a^0_{\ \,j}=0$ \cite{A43}. Then $a^0 _{\ \,0}=\sqrt{g^{00}}$ from the orthonormality of the tetrad. Let us take for ``flat" matrices  $\ \gamma ^{\sharp \, \alpha  }$ any usual Dirac matrices, for which $A=\gamma^{\sharp 0}$ [Eq. (\ref{A = gamma sharp 0})]. Thus
\be\label{a^0_j=0case}
B\equiv \,A\,\gamma^0 =\gamma^{\sharp 0}(a^0 _{\ \,0}\gamma^{\sharp 0})= \sqrt{g^{00}}\, {\bf 1}_4,
\ee
whence from (\ref{SEtilde-ES-DFW-1}):
\be\label{SEtilde-ES-DFW-3}
{\delta \mathrm{E}=-i\left[(\partial _0S)S^{-1}\right]^{a}}.
\ee
For DFW, $S(X)$ is an arbitrary ${\mathsf{Spin(1,3)}}$ transformation. Thus, $(\partial _0S)S^{-1}$ is the generic element of ${\mathcal{G}}$, the Lie algebra of ${\mathsf{Spin(1,3)}}$ -- whose the matrices 
\be
s^{\alpha \beta } \equiv [\gamma^{\sharp \alpha}  ,\gamma^{\sharp \beta }  ]\ (\alpha <\beta )
\ee 
make a basis. Hence
\be\label{deltaE}
\delta \mathrm{E}=-i \left[\omega_{\alpha \beta} s^{\alpha \beta  }\right]^a,
\ee
where, depending on the local Lorentz ${L(X)}$ that defines ${S(X)= \mathsf{S}(L(X))}$, the six coefficients ${\omega_{\alpha \beta}=-\omega_{\beta \alpha }}$ depend arbitrarily on ${X \in \mathrm{V}}$. 
With usual Dirac matrices verifying (\ref{A = gamma sharp 0}), we have 
\be 
(s^{\alpha \beta  })^\dagger =\gamma ^{\sharp 0}\, s^{\beta  \alpha }\,\gamma ^{\sharp 0},
\ee 
whence from (\ref{deltaE}) and the anticommutation formula:
\be\label{deltaE-2}
{\delta \mathrm{E}=-i\sum_{ j ,k =1}^3 \omega  _{jk } s^{jk}}.
\ee

  \section{The case with the ``chiral" Dirac matrices}

If the ``flat" Dirac matrices ${\gamma^{\sharp  \alpha} }$ are the ``chiral" ones \{see e.g. Schulten \cite{Schulten1999}, Eqs. (10.257) and (10.260)\}, we get from (\ref{deltaE-2}):
\be\label{e_jk s^jk Chiral}
{
\delta \mathrm{E}=-i\sum_{ j ,k =1}^3 \omega  _{jk } s^{jk}=\begin{pmatrix}
N  & 0\\ 
 0 &  N 
\end{pmatrix}, \qquad N\equiv -\frac{1}{2}\vec {\theta }.\vec {\sigma }
}
\ee
where ${\vec {\theta }\equiv (\theta ^k)}$ with ${\theta ^1\equiv \omega  _{23 }}$ (circular), and where ${\vec {\sigma }\equiv (\sigma  _k)}$ with ${\sigma  _k}$ the Pauli matrices. \\

Depending on the three real numbers ${\omega  _{jk}, \ 1\leq j <k \leq 3}$, the matrix ${N}$ can be {\it any} Hermitian matrix  ${2\times 2}$ with zero trace \{see e.g. Schulten \cite{Schulten1999}, Eq. (5.226)\}. Any such matrix has two eigenvalues\:  ${\mu\in {\sf R}}$ and ${-\mu }$, and has an orthonormal basis of eigenvectors: respectively ${u \in {\sf C}^2}$ for ${\mu }$, and ${v}$ for ${-\mu }$.

  \section{Non-uniqueness of the energy spectrum (DFW)}

A small perturbation of the starting Dirac matrices $\gamma ^\mu $ is defined by a similarity close to the identity: 
\be
S(\varepsilon ,X)=I+\varepsilon \, (\delta S)(X)+O(\varepsilon ^2).
\ee 
It modifies each eigenvalue of the energy operator ${\mathrm{E}}$ according to the well-known formula
\be\label{delta lambda}
\delta \lambda = (\psi \mid \delta \mathrm{E}(\varepsilon )\psi )+O(\varepsilon ^2),
\ee 
where ${\psi}$ is the eigenfunction for the unperturbed state. Inserting $\delta \mathrm{E}$ (\ref{e_jk s^jk Chiral}) and the expression (\ref{Hermitian-sigma=1-g}) of the scalar product into Eq. (\ref{delta lambda}), and decomposing the bispinor into two two-components spinors: ${\psi =(\phi ,\chi )}$, we get:
\be\label{delta lambda DFW}
{\delta \lambda = \int \psi^\dagger \delta \mathrm{E} \, \psi \ \underline{\sqrt{-g\,g^{00}}\ \dd^ 3{\bf x}}= \int(\phi ^\dagger N \phi + \chi ^\dagger N \chi)\ \underline{\dd {\sf V}}}. 
\ee
Let us fix ${\mu >0}$ and ${t}$. For any spatial position ${\bf x}\equiv (x^j)$ in our fixed chart, let us select the $2\times 2$ Hermitian matrix with zero trace, ${N=N({\bf x})}$, in such a way that ${\phi ({\bf x})}$ be the eigenvector of ${N({\bf x})}$ for the eigenvalue ${\mu}$. (The 3-vector ${\bf x}$ is the coordinate expression of the position $x$ in the space manifold $\mathrm{{M}}$ associated with the arbitrary given reference frame F \cite{A42}.) That is:
\be\label{phi croix N phi}
\phi ^\dagger N \phi = \mu\ \phi ^\dagger \phi.
\ee
Since $N$ has the eigenvalues $\mu $ and $-\mu $, we have then necessarily
\be\label{chi croix N chi}
\chi ^\dagger N \chi \geq  -\mu \ \chi ^\dagger \chi.
\ee
In Eq. (\ref{chi croix N chi}), the inequality becomes an equality only if $\chi ({\bf x})$ is an eigenvector of $N({\bf x})$ with eigenvalue $-\mu $. In that case, $\phi ({\bf x})$ is orthogonal to $\chi ({\bf x})$ in  ${\sf C}^2$, ${\chi ({\bf x})\bot \phi ({\bf x})}$. Returning to Eq. (\ref{delta lambda DFW}), we see that ${\delta \lambda >0}$ unless if i) ${\chi ({\bf x})\bot \phi ({\bf x})}$ for almost every spatial position ${\bf x}$,  {\it and}  ii) ${\int\phi ^\dagger \phi\ \dd {\sf V} =  \int \chi ^\dagger \chi\ \dd {\sf V}}$. This could occur only in extremely particular situations. In fact, i) alone implies \cite{A43} that the probability current is light-like for almost every ${\bf x}$, and this {\it is impossible} if ${m>0}$ \cite{ReiflerMorris2005}. This proves the non-uniqueness of the DFW energy spectrum, at least for a massive particle.

\section{Conclusion}

\vspace{3mm}

Three distinct gravitational Dirac equations were envisaged: the standard equation (``Dirac-Fock-Weyl" or DFW), plus two alternative equations \cite{A39} based on the tensor representation of the Dirac field (TRD). Using the hermitizing matrix ${A}$, they were studied together, to a large extent \cite{A42,A43}. Non-uniqueness results have been proved for each of the three versions \cite{A43}. In this paper, we summarized the latter work, but focussed on the results for DFW.

The Hamiltonian operator {$\mathrm{H}$} is not unique: it depends on the admissible choice of the field of Dirac matrices. The same is true for the energy operator {$\mathrm{E}$}. This is true for DFW and for TRD.

The {\it spectrum} of {$\mathrm{E}$} is itself non-unique. All of these results apply already to a {\it flat} spacetime in a non-inertial frame.

\appendix
\section{Appendix: The classical energy and its frame dependence}

One might ask if the non-uniqueness of the energy spectrum for a spin $\frac{1}{2}$ particle in a curved spacetime, reported here, might have something to do with the well-known fact that, in general relativity (GR), there is no covariant concept of local energy---already in the classical context (as opposed to the quantum context envisaged in this paper). The latter fact may be formulated by saying that there is no covariant conservation law for the (total) energy-momentum---the energy-momentum of the gravitational field giving rise only to a {\it pseudo}-tensor. As is well known, the usual ``conservation law" for the energy-momentum tensor of matter and non-gravitational fields ``does not express the conservation of anything" \cite{L&L}. However, for a classical {\it test particle,} in any arbitrary reference frame {$\mathrm{F}$}, there is a well-defined {\it Hamiltonian energy} \cite{A39,B15,Bertschinger1999}, though it depends on {$\mathrm{F}$} and, in general, on the time:

\bi
\item Geodesic motion in the Lorentzian manifold {$(\mathrm{V},g_{\mu \nu })$} derives from the (``super-")Hamiltonian over the 8-dimensional phase space associated with V: 
\be
\tilde{H}[(p_\mu),(x^\nu)]\equiv \frac{1}{2}g^{\mu \nu}((x^\rho)) p_\mu p_\nu \qquad  (c=1).
\ee 
This is because, in a (pseudo-)Riemannian manifold, geodesic motion is identical to free Hamiltonian motion, i.e., $\tilde{H}\equiv T$ with $T[(p_\mu),(x^\nu)]\equiv \frac{1}{2}g^{\mu \nu} p_\mu p_\nu$ the ``kinetic energy": Sect. 45 D in Arnold \cite{Arnold1976}. Note that {$\tilde{H}$} does not depend on the (proper) time {$\tau$}. 

\item It follows from that ``time"-independence that, in any given coordinate system, we deduce a ``normal" Hamiltonian $H$ over the usual, 6-dimensional phase space, by {\it dimensional reduction} (Sect. 45 B in Ref. \cite{Arnold1976}): 
\be
H[(p_j),(x^j),t\equiv x^0] \equiv p_0
\ee
is extracted from
\be 
g^{\mu \nu} p_\mu p_\nu -m^2 =0.
\ee
\{With the $(-+++)$ signature, we have $H \equiv -p_0$ instead, and the last equation writes $g^{\mu \nu} p_\mu p_\nu +m^2 =0$ \cite{Bertschinger1999}.\} One may note that {$E=H\equiv p_0$} is invariant under purely spatial coordinate changes (\ref{purely-spatial-change}), thus it depends only on the reference frame---which is arbitrary, but {\it fixed} in the present study.

\ei

\noindent Therefore, in a fixed reference frame, a classical test particle in a given external gravitational field has unique energy, in spite of the absence of a covariant concept of local energy in GR, where gravitational dynamics are also included in the model. Similarly, the answer to the question posed at the beginning of this Appendix is no. The non-uniqueness of the energy spectrum for a spin-half particle in a given external curved spacetime, as found in this study, has nothing to do with the absence of a covariant concept of local energy in GR.

\bigskip


\begin{thebibliography}{9}
\small

\bibitem{A43}
Arminjon M and Reifler F 2009 A non-uniqueness problem of the Dirac theory in a curved spacetime \href{http://arxiv.org/abs/gr-qc/0905.3686}{{\it Preprint} arXiv:0905.3686 (gr-qc)}

\bibitem{COW1975}
Colella R Overhauser A W and Werner S A 1975 Observation of gravitationally induced quantum interference {\it Phys. Rev. Lett.} {\bf 34} 1472--1474

\bibitem{WernerStaudenmannColella1979}
Werner S A Staudenmann J A and Colella R 1979 Effect of Earth's rotation on the quantum mechanical phase of the neutron {\it Phys. Rev. Lett.} {\bf 42} 1103--1107

\bibitem{Nesvizhevsky2002} 
Nesvizhevsky V V {\it et al.} 2002 Quantum states of neutrons in the Earth's gravitational field {\it Nature} {\bf 415} 297--299

\bibitem{BrillWheeler1957+Corr}
Brill D R and Wheeler J A 1957 Interaction of neutrinos and gravitational fields {\it Rev. Modern Phys.} {\bf 29} 465--479. Erratum: {\it Rev. Modern Phys.} {\bf 33} 623--624 (1961)

\bibitem{ChapmanLeiter1977}
Chapman T C and Leiter D L 1977 On the generally covariant Dirac equation {\it Am. J. Phys.} {\bf 44}, No. 9, 858--862.

\bibitem{A39}
Arminjon M 2008 Dirac-type equations in a gravitational field, with vector wave function {\it Found. Phys.} {\bf 38} 1020--1045 \href{http://arxiv.org/abs/gr-qc/0702048}{({\it Preprint} arXiv:gr-qc/0702048)}

\bibitem{A42}
Arminjon M and Reifler F 2008 Basic quantum mechanics for three Dirac equations in a curved spacetime \href{http://arxiv.org/abs/0807.0570}{{\it Preprint} arXiv:0807.0570 (gr-qc)}

\bibitem{B30}  Arminjon M and Reifler F 2008 Quantum mechanics for three versions of the Dirac equation in a curved spacetime \href{http://arxiv.org/abs/0810.0671}{{\it Preprint} arXiv:0810.0671 (gr-qc)}

\bibitem{Leclerc2006}
Leclerc M 2006 Hermitian Dirac Hamiltonian in the time-dependent gravitational field {\it Class. Quant. Grav.} {\bf 23} 4013--4020 \href{http://arxiv.org/abs/gr-qc/0511060}{({\it Preprint} arXiv:gr-qc/0511060)}

\bibitem{A40}
Arminjon M and Reifler F 2008 Dirac equation: Representation independence and tensor transformation {\it Braz. J. Phys.} {\bf 38} 248--258 \href{http://arxiv.org/abs/0707.1829}{({\it Preprint} arXiv:0707.1829 (quant-ph))}

\bibitem{Pauli1933}
Pauli W 1933 \"Uber die Formulierung der Naturgesetze mit f\"unf homogenen Koordinaten, Teil II: Die Diracschen Gleichungen f\"ur die Materiewellen {\it Ann. der Phys.} (5) {\bf 18} 337--354

\bibitem{Pal2007}
Pal P. B. 2007 Representation-independent manipulations with Dirac spinors \href{http://arxiv.org/abs/physics/0703214}{{\it Preprint} arXiv:physics/0703214}

\bibitem{Schulten1999}
Schulten K 1999 Relativistic quantum mechanics, in {\it Notes on Quantum Mechanics}, \href{http://www.ks.uiuc.edu/Services/Class/PHYS480/qm_PDF/chp10.pdf}{online course} of the University of Illinois at Urbana-Champaign by the same author

\bibitem{ReiflerMorris2005}
Reifler F and Morris R 2005 Hestenes' tetrad and spin connections {\it Int. J. Theor. Phys.} {\bf 44} 1307--1324 \href{http://arxiv.org/abs/0706.1258}{({\it Preprint} arXiv:0706.1258 (gr-qc))}

\bibitem{L&L} 
Landau L and Lifchitz E 1989 {\it Th\'eorie des Champs} (Moscow: Mir)\\ 
\noindent Lifshitz E M and Landau L D 1980 {\it The Classical Theory of Fields} (Oxford: Butterworth Heinemann)

\bibitem{B15} 
Arminjon M 1998 Remarks on the mathematical origin of wave mechanics and consequences for a quantum mechanics in a gravitational field {\it Proc. Sixth Int. Conf. Physical Interpretations of Relativity Theory} ed M C Duffy (Sunderland: University of Sunderland) pp 1--17 \href{http://arxiv.org/abs/gr-qc/0203104}{({\it Preprint} arXiv:gr-qc/0203104)}

\bibitem{Bertschinger1999}
Bertschinger E 1999 Hamiltonian dynamics of particle motion, in {\it General Relativity}, online course of the Massachussets Institute of Technology by the same author \href{http://web.mit.edu/edbert/GR/gr3.pdf}{web.mit.edu/edbert/GR/gr3.pdf}

\bibitem{Arnold1976}
Arnold V 1976 {\it M\'ethodes Math\'ematiques de la M\'ecanique Classique} (Moscow: Mir)\\
\noindent Arnold V I 1989 {\it Mathematical Methods of Classical Mechanics} (New York: Springer)


\end{thebibliography}
\end{document}